# Inversion-induced orbital and exchange disorders in antiferromagnetic A-site spinel CoAl₂O₄

Takashi Naka[1,*], Hiroaki Mamiya,[1] Kanji Takehana,[1] Naohito Tsujii,[1] Yasutaka Imanaka,[1] Satoshi Ishii,[2] Takayuki Nakane,[1] Minako Nakayama,[1] and Tetsuo Uchikoshi[1]

[1]*National Institute for Materials Science, 1-2-1 Sengen*

*Tsukuba, Ibaraki 305-0047, Japan*

[2]*Department of Physics, Tokyo Denki University, Hatoyama, Saitama 350-0394, Japan*

(Received                    )

The inversion and volume effects on magnetism in a spinel-type magnetically frustrated compound, CoAl₂O₄, and its gallium-substituted system, CoAl$_{2-x}$Ga$_x$O₄, were investigated. Magnetically frustrated Co$^{2+}$ with spin $S = 3/2$ on the tetrahedral site formed a diamond lattice in CoAl₂O₄ located in the vicinity of the magnetic phase boundary between Néel and spin-spiral states. In the Ga-substituted system, the number of Co ions, the so-called inversion η dominating the octahedral site, increased with increasing $x$. From comprehensive crystallographic, magnetic, and thermal measurements, increments of both volume and inversion strongly reduced the Néel point, while the latter also induced a spin-glass state above the critical value of $η_c = 0.09$. In the spin glass state, $η > η_c$, the orbital degree of freedom of Co$^{2+}$ ions in the octahedral site appeared in the magnetic entropy, which couples strongly with that of spin, even above the magnetic transitions. Above $η \sim η_c$, the field-induced quenched magnetic moment appeared above the transitions. Therefore, a short range ordered state emerged among the paramagnetic, antiferromganetic, and spin-glass states in the magnetic phase diagram.

## I. INTRODUCTION

Magnetic frustration caused by competition between interactions results in suppression of the magnetic-ordering temperature and large pseudo-degeneracies of the ground state. This leads to large residual entropy remaining at relatively low temperatures. Consequently, not only do exotic magnetic states (e.g., quantum and classical spin-liquid and spin-spiral states) arise, but also the magnetic ground state is quite sensitive with respect to physical and chemical perturbations (e.g., magnetic field and pressure and crystallographic disorder in magnetically frustrated systems). In an A-site spinel magnet (diamond lattice magnet), which could be considered the three-dimensional counterpart of the two-dimensional honeycomb antiferromagnet, a Néel state (magnetically bipartite) is realized when the nearest neighbor interaction $J_1$ acting between spins located at the A-site of the spinel structure is considered. The diamond lattice consists of two interpenetrating, magnetically frustrated fcc lattices shifted along the [111] direction. However, quite often in A-site antiferromagnets the magnetic transition temperature $T_N$ is suppressed strongly compared with the paramagnetic Curie temperature θ subtracted from the temperature dependence of the magnetic susceptibility. This is expressed by the so-called frustration parameter $f = |\theta|/T_N \gg 1$. Although an antiferromagnetic state can be stabilized by a nearest-neighbor interaction, if a next-neighbor interaction $J_2$ is introduced, frustration effects are anticipated. Note that cobalt aluminate $CoAl_2O_4$ can be a highly frustrated spin system because it shows a large suppression of $T_N$ ($f \sim$ 10). In a Monte Carlo simulation derived temperature–($J_2/J_1$) magnetic phase diagram [1], the magnetic phase boundary between the Néel and spin-spiral states is at $J_2/J_1 = 1/8$. The magnetic state of $CoAl_2O_4$ might be expected to be situated in the vicinity of the phase boundary. However, experimentally, it is controversial whether the magnetic ground state is a Néel (AF), spin liquid (SL), or spin-spiral (SS) state [2–9]. One reason the experimental controversy has persisted for a few decades is that inversion of $CoAl_2O_4$ specimens has been unavoidable. This is because the inversion, which indicates the level of defect of the crystallographic ionic configuration for the spinel structure, seriously affects the magnetic ground state of $CoAl_2O_4$. More recently, by studying the linear magnetoelectric effect and using a Monte Carlo simulation, Ghare *et al*. concluded that $CoAl_2O_4$ with low inversions is a collinear antiferromagnet that exhibits a spin glass transition with higher inversions [9].

It has been observed that the $T_N$ of $CoAl_2O_4$ is enhanced by applying hydrostatic pressure [10]. This suggests that the volume can be a control parameter, so it might be possible to tune the magnetic state of $CoAl_2O_4$ from the Néel to a spin-spiral state beyond the boundary of $J_2/J_1$ = 1/8. Motivated by this speculation, Melot *et al*. generated the magnetic phase diagram of $CoAl_{2-x}Ga_xO_4$, the volume of which increases with increasing $x$ [11]. It is quite difficult, however, to distinguish the volume and inversion effects on the magnetic behavior of $CoAl_{2-x}Ga_xO_4$, because both the inversion and the volume expansion suppress the AF state and, therefore, create a spin-glass state at $\eta > \eta_c$ [11]. In this paper, we describe the effect of inversion on the magnetic behavior and specific heat, particularly around the magnetic phase boundary at $\eta \sim \eta_c$. Combined with previous studies [4, 10–11], we attempt to distinguish both effects. The Ga-substituted system is indeed suitable for this purpose because, as is shown below, the inversion can be controlled in the range $0.055 \leq \eta \leq 0.66$ for samples synthesized in this work. In a previous study [4], while the spin glass (SG) state emerged above $\eta \sim 0.08$ in quenched $CoAl_2O_4$ samples, it was not yet clear how and why the antiferromagnetism was destroyed and a SG state emerged. One scenario is that around the AF-SS phase boundary, the magnetic ground state is particularly sensitive to disorder. Although the competition and interplay between magnetic frustration and disorder that appear in many frustrated spin systems are ubiquitous, a comprehensive description and predictions regarding the disorder effects on spin liquids and frustrated spin systems are needed. Despite previous study from this point of view [11], we could overcome this new aspect, that is, the macroscopic magnetic evolution in $CoAl_{2-x}Ga_xO_4$, by varying the inversion and volume. Here we propose an inversion–temperature phase diagram in the vicinity of the phase boundary.

## II. EXPERIMENTAL

Polycrystalline Ga-substituted samples $CoAl_{2-x}Ga_xO_4$ and $ZnAl_{2-x}Ga_xO_4$ with $0 < x \leq 2$ were prepared by solid–solid reaction of the proper amounts of CoO (3N), $Ga_2O_3$ (4N), $Al_2O_3$ (4N), and ZnO (3N). The latter was used as a reference sample to evaluate the lattice contribution to specific heat. The mixed powder was heated at 1300 °C for 24 h and then cooled to room temperature at a rate of 36 °C/h. To estimate precisely the cation distribution, we conducted powder x-ray diffraction measurements using a synchrotron x-ray source and made crystal structure refinements with the Rietveld method. Powder x-ray diffraction measurements using synchrotron radiation ($\lambda = 0.65296$ Å) were conducted on the BL15XU beam line at SPring-8 (Harima, Japan) [12]. The diffraction data revealed no secondary oxide phase such as CoO, $Co_3O_4$, or $\alpha$-$Al_2O_3$ in any of the samples. The crystal structure of the samples was refined with the Rietveld method by using the RIETAN-FP program [13]. Dc-magnetization and specific heat down to $T = 2$ K were measured by using a conventional superconducting quantum interference device (SQUID) magnetometer (MPMS-XL; Quantum Design) and a physical property measurement system (PPMS; Quantum Design), respectively. For high-field magnetization measurements at $T = 1.75$ K and $B = \pm 15$ T, we employed a high-field magnetometer combined with the extraction method in the Tsukuba Magnet Laboratory, National Institute for Materials Science (NIMS). Thermoremanent magnetization (TRM) was measured with a SQUID magnetometer at $H = 0$ after cooling from higher temperatures ($T >> T_N$ or $T_{sg}$) under a magnetic field of strength $H_{FC} = 100$ Oe at $T = 2$ K for a duration of 200 s.

### III. RESULTS AND DISCUSSION

#### 1. Crystallographic parameters

Figure 1(a) shows the $x$-variations of lattice constant $a$ and oxygen positional parameter $u$ (so-called $u$-parameter) for $CoAl_{2-x}Ga_xO_4$. As expected, the lattice constant increased with increasing $x$, while $u$ decreased. The octahedron of $BO_6$ (the B-site is surrounded octahedrally by six oxygen ions) was distorted, and the position of the oxygen ions deviated from that of the regular octahedron of $u = 0.25$. It seems that not only $a$ (volume) but also $u$ affected the magnetic interactions between spins localized on $Co^{2+}$. The $x$-variations of $a$ and $u$ agreed well with those reported by Melot *et al.* [11]. As previously pointed out [8], the magnetic state in a diamond lattice magnet is influenced strongly by inversion. The Ga-substituted system is known to exhibit larger inversion because the $Ga^{3+}$ ion prefers to distribute into both the A- and B-ions [11]. The A-site occupancies of Co, Al, and Ga, denoted as $g\_CoA$, $g\_AlA$, and $g\_GaA$, respectively, are shown in Fig. 1(b) as a function of $x$. Note that $\eta = 1 - g\_CoA = g\_AlA + g\_GaA$. The inversion increased linearly with increasing $x$ and saturated at $\eta \sim 0.6$ above $x \sim 1.4$. In the end number $CoGa_2O_4$ with $\eta = 0.66$, the Ga ion distributed nearly randomly into the A- and B-sites. For the reference samples $ZnAl_{2-x}Ga_xO_4$, $a$ and $u$ are plotted in Fig. 1(a) as a function of $x$. Because the atomic numbers of Zn and Ga are quite close ($A_{Zn} = 30$ and $A_{Ga} = 31$), the site occupancies were difficult to obtain appropriately by refinement, so we assumed $\eta$ to be zero ($g\_ZnB = 0$) for all values of $x$.

## *2. Magnetization*

The magnetic state changed from antiferromagnetic-like to spin glass with increasing η (Fig. 2(a)), as indicated previously [11]. The magnetic transition point could be well defined as the sharp maximum point of $d\chi(T)/dT$ (Fig. 2(b)), hereafter denoted $T_{max}(d\chi/dT)$, which coincided quite well with the maximum point of the specific heat divided by temperature, $T_{max}(C/T)$, in $CoAl_2O_4$ [10]. At the magnetic boundary, the shapes of both $\chi(T)$ and $d\chi(T)/dT$ changed drastically. As typically observed in SG systems, $\chi(T)$ measured after zero-field cooling (ZFC) exhibited a cusp at $T = T_{sg}$, and the irreversibility between susceptibilities measured after ZFC and field cooling (FC) appeared below $T_{sg}$. In the AF state, $\chi(T)$ had a rather broad maximum slightly above $T_N$. The effective magnetic moment $p_{eff}$ was nearly constant with $x$, whereas the Weiss temperature increased with increasing $x$ (Fig. 2(c)). The magnetic boundary seemed to be located at $x = 0.09$ (Fig. 2(d)). As seen in $Co_{1-x}Zn_xAl_2O_4$ [10], a field-induced quenched magnetic moment $m_0$ was observed in $CoAl_{2-x}Ga_xO_4$ (Fig. 3). This is defined by $m_0 = (\chi(H_{FC}) - \chi(H_{FC} = 0))H$. Interestingly, $m_0$ increased with increasing $x$ and exhibited a small anomaly at the magnetic phase boundary. As mentioned below, $m_0$ correlated strongly with η and could be a measure of inversion. Indeed, $m_0$ enhanced above the magnetic phase boundary between the AF and SG phases followed a power law with a threshold, $m_0 = m_c(\eta - \eta_{mc})^\alpha$, as shown in Fig. 3, where $\eta_{mc} = 0.087$ and $\alpha = 0.87$ were obtained by a least squares fit. As mentioned above, the threshold $\eta_{mc}$ agreed well with the phase boundary $\eta_c$.

## *3. Specific heat and magnetic entropy*

Correspondingly, a magnetic anomaly could be detected by specific heat. It was previously reported that a maximum in $C(T)/T$ for $CoAl_2O_4$ with $\eta = 0.055$ was apparent at $T_{max}(C/T) = 8.5 \pm 0.4$ K at ambient pressure [10]. With increasing $x$ in $CoAl_{2-x}Ga_xO_4$, $T_{max}(C/T)$ decreased and reached a minimum at $x = 0.1$–0.2, while the peak became broader (Fig. 4). Figures 5(a)–(d) show the temperature dependence of $C/T$ for $CoAl_{2-x}Ga_xO_4$ for $x = 0$, 0.6, 1.2, and 2, respectively. In addition, $C(T)/T$ for the reference materials $ZnAl_{2-x}Ga_xO_4$ is represented in the respective figures. The magnetic contribution to the specific heat can be calculated by subtraction of the lattice contribution of $ZnAl_{2-x}Ga_xO_4$, that is, $\Delta C/T = [C(CoAl_{2-x}Ga_xO_4) - C(ZnAl_{2-x}Ga_xO_4)]/T$. The magnetic entropy $S_{mag}$, which can be obtained by an integration of $\Delta C/T$ with respect to temperature, is also depicted in the respective figures. At $x = 0$, $S_{mag}(T)$ saturated to a value of $S_{mag}/R\ln(2S+1) = 92\%$ above temperatures comparable with the Weiss temperature, $T \sim |\theta|$, where $S = 3/2$ for $Co^{2+}$ and $R$ is the gas constant. In the Ga-substituted system of $x = 0.6$, 1.2, and 2, a broad maximum in $C/T$ was observed at temperatures somewhat higher than $T_{sg}$. Remarkably, $S_{mag}$ for $x \geq 0.6$ did not show a saturation below 100 K and seemed to exceed the spin contribution $R\ln(2S+1)$ at higher temperatures. Indeed, an additional contribution seemed to result from an orbital degree of freedom of the electronic state of $Co^{2+}$ located at the octahedral B-site. The low-lying 3d-state dε is threefold degenerate and was occupied by five electrons. The threefold orbital degeneracy of dε was expected for the $Co^{2+}$ ion at the B-site. Therefore, $S_{mag}(x)$ seemed to be represented by $S_{mag} = R\ln(2S+1) + \eta(x)R\ln(2\tau + 1)$, where the first and second terms are the spin and orbital contributions to the magnetic entropy, respectively, and $\tau$ is the pseudo-orbital moment for $Co^{2+}$ located at the B-site. Note that the threefold degenerate dε state stabilized by the octahedral ligand field provided an orbital degree of freedom $\tau = 1$, while the low-lying dγ state stabilized in the tetrahedral ligand field has no orbital degree of freedom $\tau = 0$. Compared with the value of $S_{mag}$ for $x = 0$, $S_{mag}(T = |\theta|)$ at $x \geq 0.6$ was remarkably suppressed to 70–60% of the spin contribution $R\ln(2S+1)$. This suggests that a strong coupling between $Co^{2+}$ spins located at the A- and B-sites and/or a spin-orbital entanglement of $Co^{2+}$ at the octahedral B-site was realized at $T > T_{sg}$.

At high temperatures, $S_{mag}$ saturated as expected to $R(\ln 4 + \eta(x)\ln 3)$. It is plausible that, in addition to the exchange interaction represented by $\theta$, a characteristic energy scale for a spin-orbital entanglement in $CoAl_{2-x}Ga_xO_4$ was provided. Here we denote this energy scale to be $\Theta$. As mentioned previously, the magnetic entropy evolution in $FeSc_2S_4$ is related to the crystallographic (an orbital ordering) and magnetic (an antiferromagnetic ordering) transitions recently reported [14]. With increasing $\eta$, fluctuations associated with a spin-orbit entanglement develop and contribute to the destruction of the long range AF ordering. It is well known that $Co^{2+}$ in octahedral ligand fields is weakly Jahn–Teller active, leading to a crystallographic structural transition and/or instability. In $CoAl_{2-x}Ga_xO_4$, a co-operative Jahn–Teller transition seemed not to occur, because the B-site occupancy of Co, $g\_CoB \sim 0.03$–$0.33$, was less than the bond percolation threshold $0.39$ for the B-site [15]. Nevertheless, incoherent (local) distortion of the $CoO_6$ octahedron was expected below $T \sim \Theta$, which resulted in exchange disorder. The spin arrangement is modified by the exchange disorder when $|\theta| > \Theta$ and the long-range antiferromagnetic ordering is destroyed by the orbital disorder at $|\theta| \sim \Theta$ and emergence of the SG state at $|\theta| < \Theta$. In $CoAl_2O_4$, the antiferromagnetic domain observed in neutron diffraction has several sizes of lattice constants and does not grow macroscopically, even at $T << T_N$ [5]. The first-order characteristics of the domain formation observed in neutron scattering and the field-quenched magnetization $m_0$ indicated in magnetism are understandable if one considers that the orbital disorder is quenched gradually under the conditions of $|\theta|$, $\Theta$ $>> T_N$ or $T_{sg}$.

These findings and the large frustration parameter suggest that there is a short range AF cluster nucleated by the orbital/exchange disorder in the spin liquid state at low temperatures $T << |\theta|$, $\Theta$. We can evoke a scenario in which a magnetic cluster is nucleated, preferably around the quenched disorders, bringing about magnetic incoherency in the lattice. The short-range magnetic ordering resulted from the inversion in the A-site spinel magnet because an inverted $Co^{2+}$ spin in the B-site created a strong magnetic interaction between $Co^{2+}$ spins in the A-site via the super exchange interaction A-O-$Co^{2+}$(B)-O-A [2, 16].

The SG-like behaviors observed in antiferromagnetic quasicrystal rare-earth compounds, such as the icosahedral R-Mg-Zn (R = rare-earth elements), are interpreted to be located between those of canonical spin glass and superparamagnetic materials [17–18]. In icosahedral Ho-Mg-Zn, neutron diffraction using a single-quasicrystalline sample has revealed that a short-range correlation (SRC) between $Ho^{3+}$ moments develops below $T = 5$ K [19]. The observed SRC seems to be incompatible with the highly degenerate free energy surface characteristic of canonical spin glasses [18]. The fragmentation of the magnetic domain structure inferred to be at the AF–SG boundary seems to result in the observed SG-like behavior. In microscopic magnetic investigation, such as by neutron diffraction, nuclear magnetic resonance, and muon spin resonance spectroscopy, of $CoAl_2O_4$ polycrystal [8, 20] and single crystal specimens [5, 7, 21–23], the spin system does not show any long-range antiferromagnetic orderings, even well below the magnetic transition, and retains its nano-sized antiferromagnetic domain structure [5]. MacDougall *et al*. [5, 22] have suggested that magnetic fragmentation occurs as a result of a kinetically inhibited magnetic order that might be associated with a first-order phase transition. These observations are quite compatible with those of the above-mentioned quasicrystal antiferromagnetic materials. In both spin systems, magnetic frustration and high sensitivity with respect to magnetic and crystallographic disorders seem to be essential for understanding their anomalous magnetic behaviors. For example, the emergence of the field-quenched moment and the discrepancy between magnetizations after ZFC and FC seem to be ubiquitous in A-site spinel magnets.

## 4. Field-quenched moment and η–T phase diagrams

Following the experimental determination of $T_N$ [10], the $x$-variation of the maximum temperatures for $d\chi/dT$ and $C/T$, namely $T_{max}(d\chi/dT)$ and $T_{max}(C/T)$, are plotted in Fig. 2(d). The distributions correspond well with the $x$–$T$ phase diagram reported previously by Melot *et al.* [11]. Figure 6 shows the η–$T$ phase diagram for CoAl$_{2-x}$Ga$_x$O$_4$. The tri-critical point at ($\eta_{tc}$, $T_{tc}$) = (0.09, 4.1 K) is where the three magnetic phases, paramagnetic (P), AF, and SG, meet. The AF–SG phase boundary $\eta_c = 0.09$ for CoAl$_{2-x}$Ga$_x$O$_4$ agreed with the $\eta_c$ of ~ 0.08 observed for CoAl$_2$O$_4$ [4]. We could deduce, therefore, that below $\eta = \eta_c$ the decrement ratio of $T_N$ with respect to η, $\Delta T_N/\Delta\eta$, was somewhat greater than that with respect to volume, $\Delta T_N/\Delta V$ for CoAl$_{2-x}$Ga$_x$O$_4$, as follows. The volume derivative of $T_N$ was estimated to be the rather large value of $(\partial \ln T_N/\partial \ln V)_\eta = -8.8$ under the condition of $\eta$ = constant [4]. Consequently, from the linear variation of $T_N$ with respect to η below $\eta_c = 0.09$ in CoAl$_{2-x}$Ga$_x$O$_4$, the η-derivative of $T_N$ $(\partial \ln T_N/\partial \ln \eta)_V = -0.79$ could be estimated. This value was somewhat larger, but comparable, with $-0.3 \pm 0.1$ for CoAl$_2$O$_4$, which was subtracted from the data of $T_N(\eta)$ reported previously [4]. Note that for CoAl$_2$O$_4$ the volume was constant with respect to η. The contributions of $(\Delta T_N(\eta))_V$ and $(\Delta T_N(V))_\eta$ and to the total decrement of $T_N$, $(\Delta T_N)_{tot} = (\Delta T_N(\eta))_V + (\Delta T_N(V))_\eta$, were 0.95 and 0.05, respectively, below $\eta_c$ for CoAl$_{2-x}$Ga$_x$O$_4$. Although the $x$-variation of $T_{sg}$ could be seen, the contributions of both $\Delta T_{sg}(\eta)$ and $\Delta T_{sg}(V)$ could not be distinguished quantitatively. The spin glass temperature $T_{sg}$ increased with increasing $x$ for CoAl$_{2-x}$Ga$_x$O$_4$.

Figure 7 shows the magnetization (MH) curves measured at $T = 1.7$ K for $x = 0$, 0.2, and 0.6 for CoAl$_{2-x}$Ga$_x$O$_4$. In the Ga-substituted system, the MH curves obviously exhibited hysteresis, as indicated previously [11]. For $x = 0.6$, the coercive field was $H_c = 2.5$ kOe, and the remanent magnetization was $m_r = 0.016$ μ$_B$, comparable with $m_0 = 9.6 \times 10^{-3}$ μ$_B$ obtained at $T = 2$ K. To indicate the magnetic response corresponding with the magnetic entropy released above the magnetic transition, mostly at $\eta \gg \eta_c$, we measured the temperature dependence of TRM, $M_{TR}(T)$ (Fig. 8(a)). Above $x = 0.04$, the $dM_{TR}(T)/dT$ curve exhibited two anomalous steps at $T_H$ and $T_L$, while $T_L(x)$ agreed well with $T_{sg}(x)$, as shown in Fig. 8(b). Although at $x = 0$ $T_H$ corresponded with $T_{max}(d\chi/dT)$ and $T_{max}(C/T)$, with increasing $\eta$ $T_H$ deviated in the direction above $T_{max}(d\chi/dT)$ and approached $T_{sg}$ gradually above $\eta = \eta_c$. From the neutron diffraction measurement for a CoAl$_2$O$_4$ specimen with higher inversion $\eta \sim \eta_c$, it was found that a short range antiferromagnetic correlation strongly developed above $T_N$. Consequently, the temperature region $T_L < T < T_H$ for CoAl$_{2-x}$Ga$_x$O$_4$ could be assigned to the short-range ordered (SRO) state. It is noteworthy that the maximum temperature of $C/T$ (see Fig. 4) was quite close to $T_H$ (Fig. 6). These facts show that the SRO arising at $T > T_N$ or $T_{sg}$ was a bulk property in cobalt spinel with certain degree of inversion, but not the result of a tiny amount of any impurity phase that was undetectable by XRD with a high flux synchrotron beam.

## IV. CONCLUSION

We reported the effects of Ga substitution in A-site cobalt spinel on the crystal structure, magnetization, and specific heat, and proposed composition–temperature ($x$–$T$) and inversion–temperature ($\eta$–$T$) magnetic phase diagrams for CoAl$_{2-x}$Ga$_x$O$_4$ in the range $0 \leq x \leq 2$. The inversion $\eta$ was controlled in the range of $0.055 < \eta < 0.66$, whereas the lattice volume expanded linearly with increasing $x$. The field-induced ferromagnetic component $m_0$ was enhanced above the magnetic phase boundary between the AF and SG phases ($\eta_c = 0.09$) and followed a power law, $m_0 = m_c(\eta - \eta_{mc})^\alpha$ with $\eta_{mc} = 0.087$ and $\alpha = 0.87$. Because of the additional orbital degree of freedom of the Co$^{2+}$ at the B-site, the excess magnetic entropy seemed to be well above the magnetic transition temperatures. The strong correlation between the spin and orbital degrees of freedom and the strong exchange coupling via Co$^{2+}$-spin on the B-site might have resulted in suppression of the Néel state (long range antiferromagnetic ordering) and brought about spin-glass and short range ordered states. Above the magnetic and spin-glass transitions, a short range magnetic ordered state emerged at $T_H$, where thermoremanent magnetization and also an anomalous peak in specific heat ($C/T$) appeared.

## ACKNOWLEDGEMENTS

We appreciate the help of the Material Analysis Station of the National Institute for Materials Science for the inductively coupled plasma–mass spectrometry analyses. This work was partially supported by a Grant-in-Aid for Scientific Research, KAKENHI, (25246013, 16K13999). The synchrotron radiation experiments were performed on the BL15XU beam line at SPring-8 with the approval of the Japan Synchrotron Radiation Research Institute (2015B4505, 2016B4505, 2018A4507).

**Figures and figure captions**

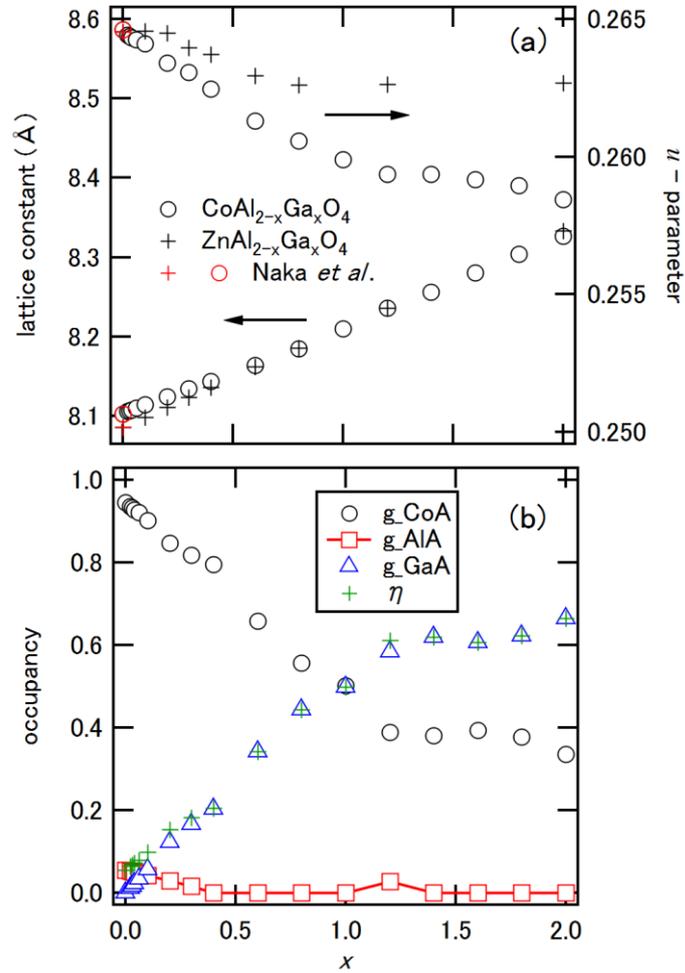

FIG. 1. (a) The $x$-variations of the lattice constant and $u$-parameter for $CoAl_{2-x}Ga_xO_4$ (open circles) and $ZnAl_{2-x}Ga_xO_4$ (crosses) for $0 < x \leq 2.0$ at room temperature. Red circles and crosses represent the lattice constant and $u$-parameter for $CoAl_2O_4$ and $ZnAl_2O_4$, respectively, obtained previously [10]. The error bars on the lattice constant points are smaller than the symbols. (b) The A-site occupancies $g\_CoA$, $g\_AlA$, and $g\_GaA$ of the cations of Co, Al, and Ga, respectively, and inversion η as a function of $x$.

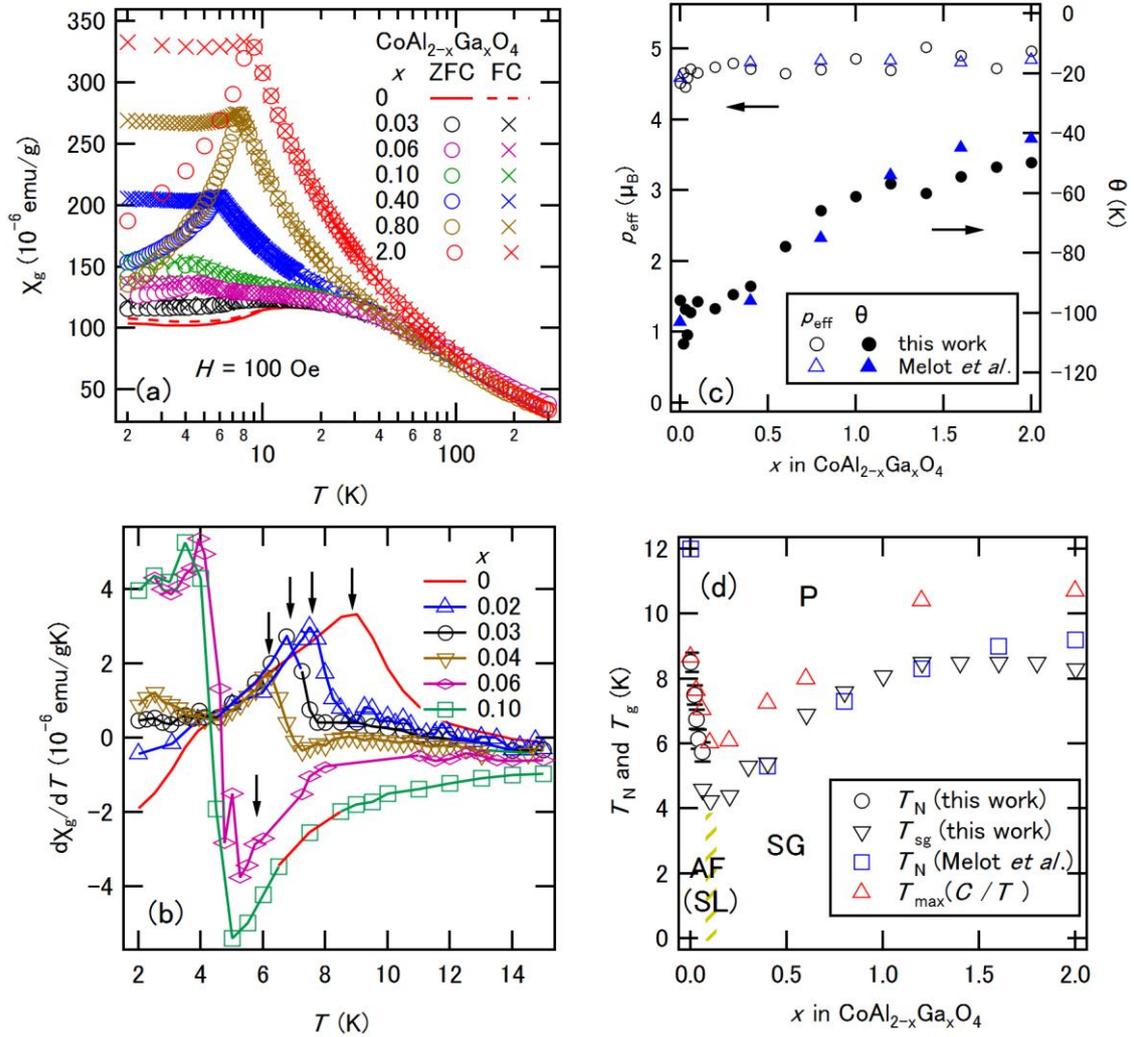

FIG. 2. (a) Magnetic susceptibility after ZFC and FC for CoAl$_{2-x}$Ga$_x$O$_4$ as a function of temperature measured at $H = 0.1$ kOe. (b) Temperature derivative of the susceptibility as a function of temperature. Arrows indicate the magnetic transition temperature $T_N$. (c) Effective magnetic moment $p_{eff}$ and Weiss temperature θ as functions of $x$. (d) Composition–temperature phase diagram. The hatched area around $x = 0.06$ represents the magnetic phase boundary between antiferromagnetic (AF) and spin glass (SG) states. P and SL denote paramagnetic and spin liquid states, respectively.

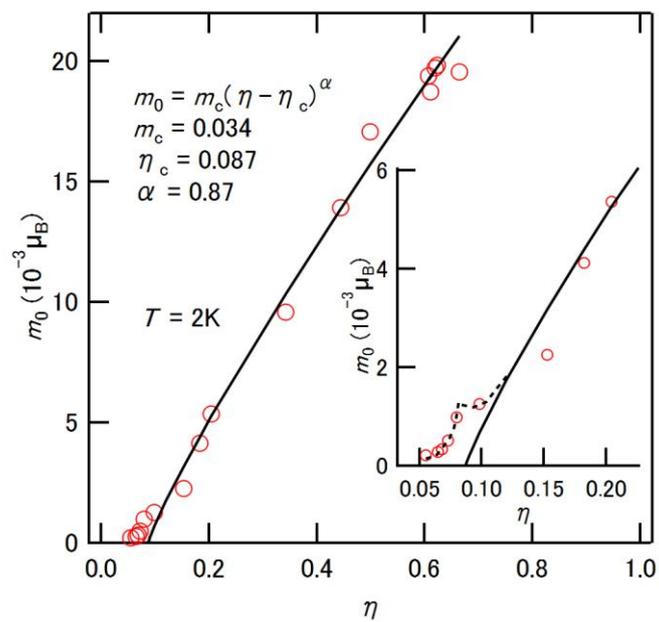

FIG. 3. η-variation of the quenched ferromagnetic component $m_0$ at $T = 2$ K. The solid line represents the fitted curve of $m_0 = m_c(\eta - \eta_c)^\alpha$. The inset shows an expanded view of the lower η region. The dashed curve is an aid to visualization.

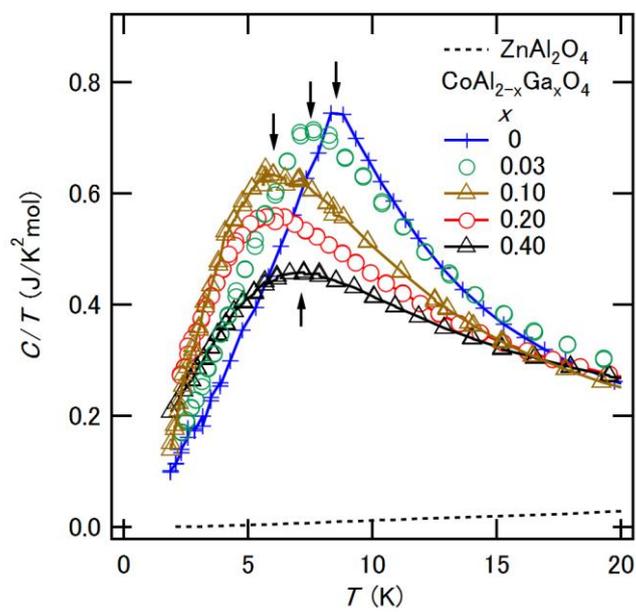

FIG. 4. Temperature dependence of $C/T$ for $CoAl_{2-x}Ga_xO_4$ below $x = 0.4$. Vertical arrows indicate the maximum temperature $T_{max}(C/T)$.

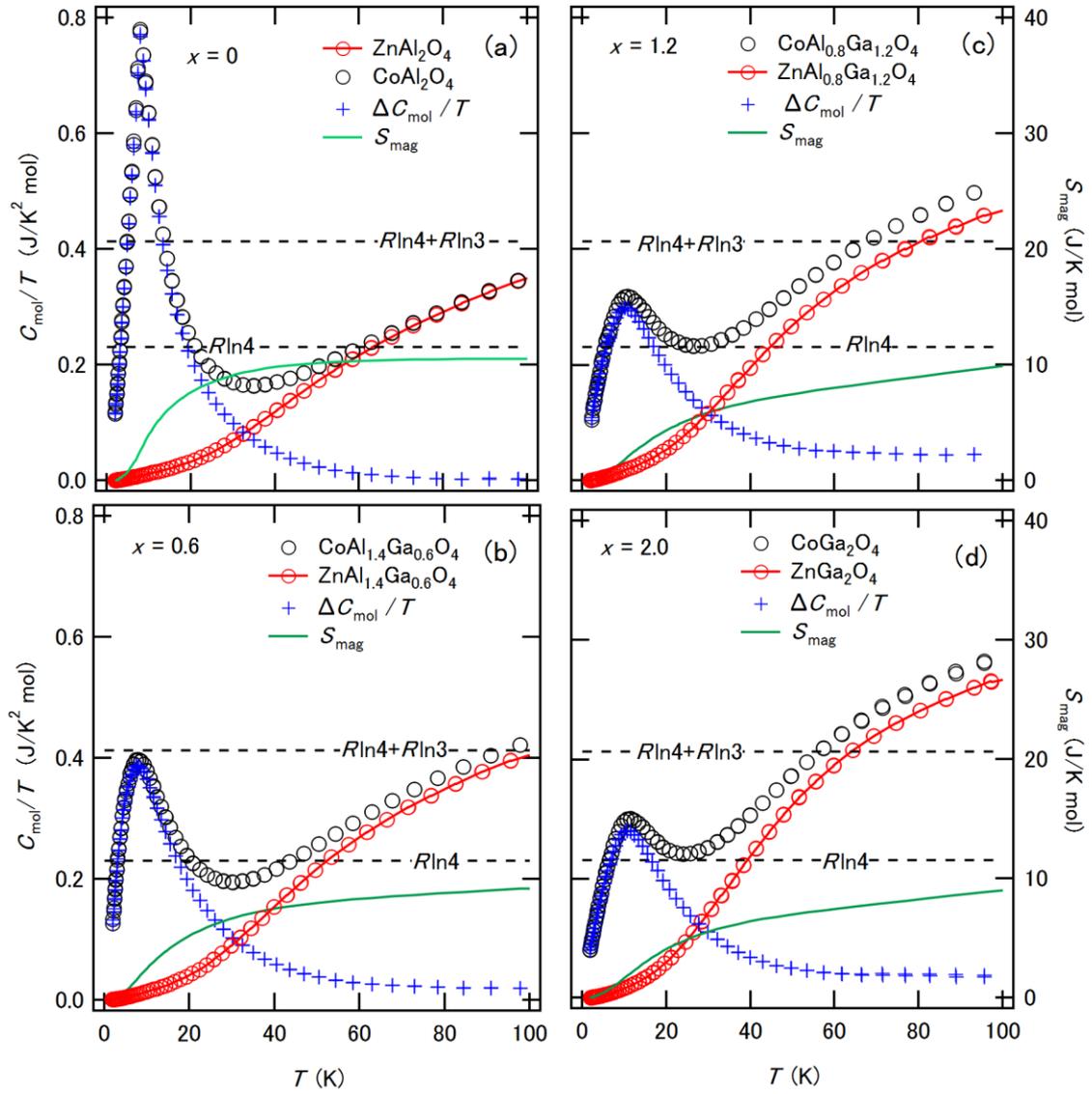

FIG. 5. The temperature dependences of $C_{mol}/T$ for $CoAl_{2-x}Ga_xO_4$ and $ZnAl_{2-x}Ga_xO_4$ (lattice contribution) for (a) $x = 0$, (b) $x = 0.60$, (c) $x = 1.2$, and (d) $x = 2.0$. The blue crosses and green solid lines represent the magnetic specific heat $\Delta C_{mol}/T$ and magnetic entropy $S_{mag}$, respectively.

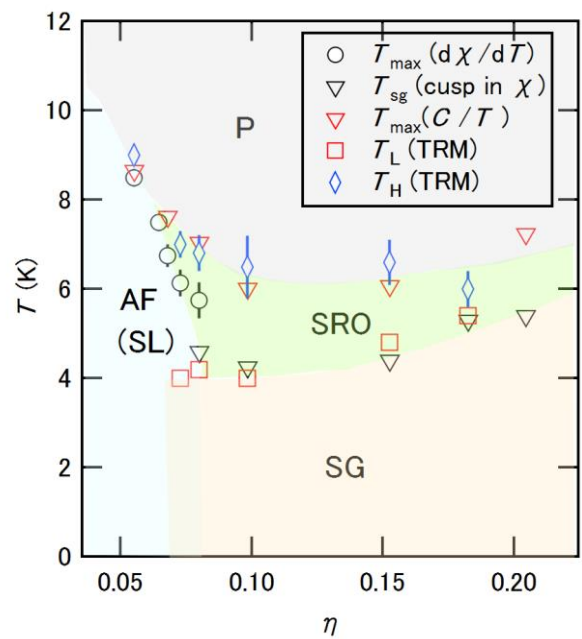

FIG. 6 The η–T phase diagram for CoAl$_{2-x}$Ga$_x$O$_4$. P, AF, SL, SRO, and SG denote the paramagnetic, antiferromagnetic, spin-liquid, short-range ordered, and spin glass states, respectively.

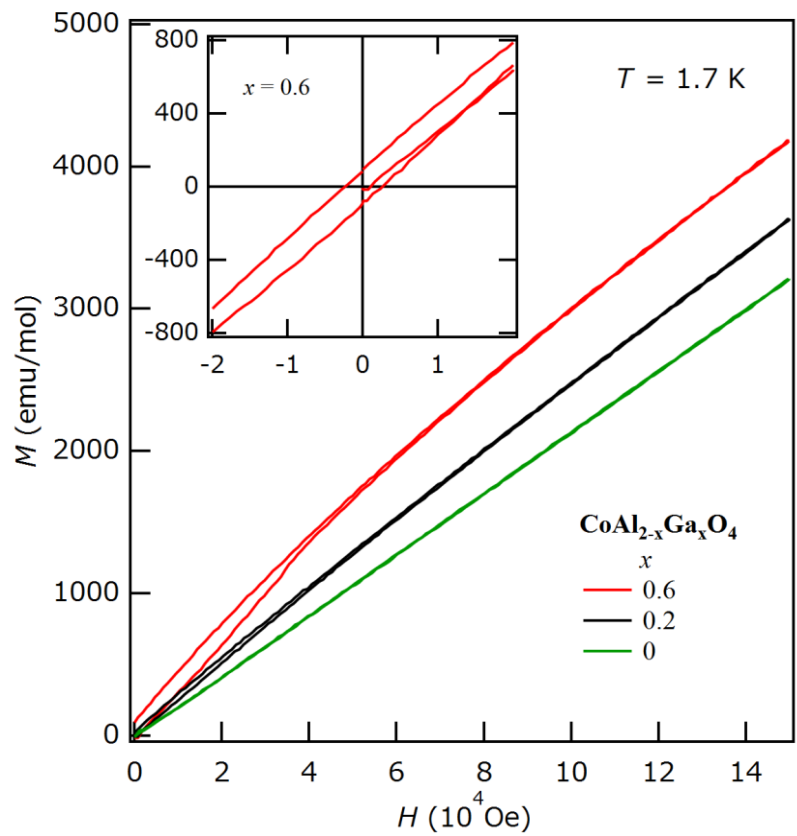

FIG. 7 Magnetization (MH) curves up to $\mu_0 H = 15$ T for CoAl$_{2-x}$Ga$_x$O$_4$ measured at $T = 1.7$ K. Inset shows an expanded view around $H = 0$ for CoAl$_{1.4}$Ga$_{0.6}$O$_4$.

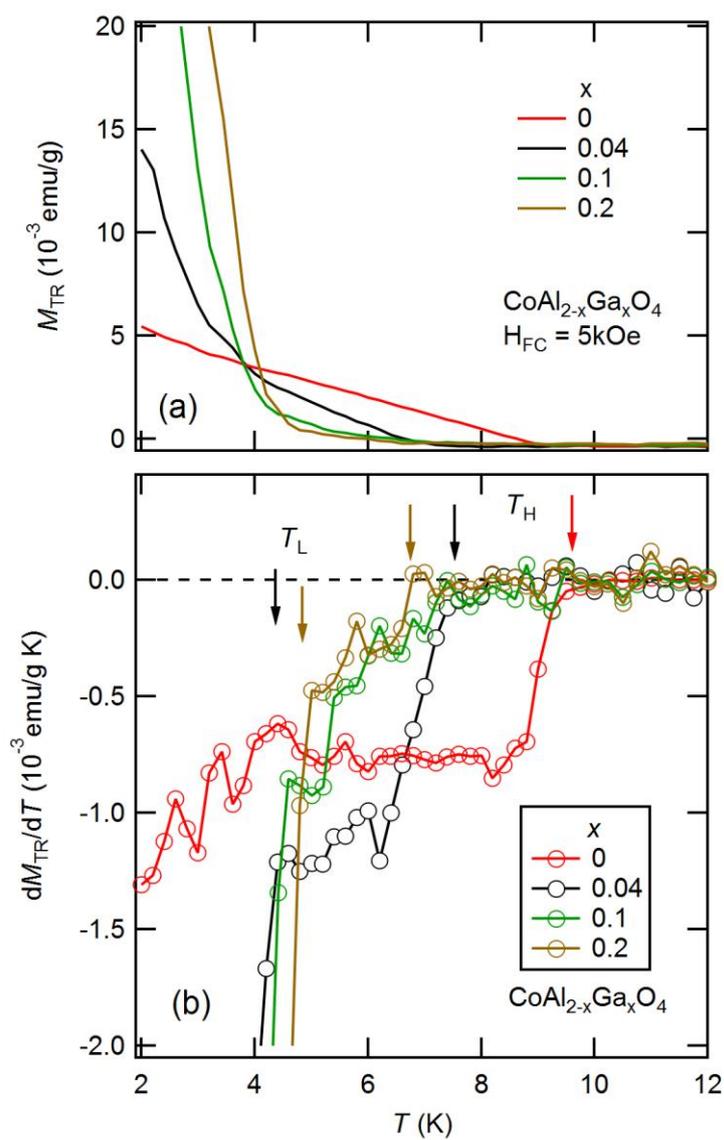

FIG. 8 The temperature-variations of (a) thermoremanent magnetization $M_{TR}$ and (b) the derivative of $M_{TR}$

with respect to temperature for $x = 0$, 0.04, 0.1, 0.2, and 2.0. The arrows indicate the anomalous points denoted as $T_H$ and $T_L$.